# Controlling Multimode Optomechanical Interactions via Interference

Mark C. Kuzyk and Hailin Wang

Department of Physics, University of Oregon, Eugene, OR 97403, USA

Abstract

We demonstrate optomechanical interference in a multimode system, in which an optical mode couples to two mechanical modes. A phase-dependent excitation-coupling approach is developed, which enables the observation of constructive and destructive optomechanical interferences. The destructive interference prevents the coupling of the mechanical system to the optical mode, suppressing optically-induced mechanical damping. These studies establish optomechanical interference as an essential tool for controlling the interactions between light and mechanical oscillators.

Cavity optomechanics explores fundamental interactions between light and mechanical oscillators[1,2]. While earlier research efforts have focused on simple two-mode systems, in which an optical mode couples only to one mechanical mode, recent efforts have also emphasized multimode systems, in which an optical (or mechanical) mode couples to multiple mechanical (or optical) modes. These multimode systems can enable and provide a versatile experimental platform for a rich variety of physical phenomena, such as exceptional points and topological energy transfer[3], back-action evasion[4,5], two-mode squeezing[6,7], and optical or mechanical state transfers[8-18].

Interference plays a pivotal role in quantum control of multi-level or multi-qubit systems. The advances on multimode systems have thus stimulated strong interest in exploring optomechanical interference processes and in using these processes for applications such as optomechanically-mediated interfaces, entanglement, and ground state cooling[19-24]. For example, when two mechanical modes couple to a common optical mode[3,5-7,13,17,25-27], destructive interference between the respective optomechanical processes can prevent the coupling of the mechanical system to the optical mode, leading to the formation of an optically-dark mechanical superposition mode[23,26]. Similarly, a mechanically-dark optical superposition mode can be formed when two optical modes couple to a common mechanical mode[19,20]. These dark modes can be used for the realization of state transfer as well as two-mode squeezing. The dark optical mode can also be exploited to circumvent the effects of thermal mechanical noise[19,20,22,23]. Evidence for dark optical and dark mechanical modes has been reported in earlier studies[8,26], though there has been no direct experimental probe on the underlying optomechanical interference process.

In this paper, we report experimental demonstration and manipulation of optomechanical interference in a multimode system, in which an optical mode couples to two mechanical modes. We observe constructive and destructive optomechanical interferences by using a phase-dependent excitation-coupling approach and by varying the relative phase of the optical fields that couple to the respective mechanical modes. With a phase shift of π, these interference processes can effectively switch the mechanical system from an optically-active to an optically-dark superposition mode. Additional experiments on the decay of the dark mode demonstrate directly the suppression of optically-induced mechanical damping and thus the decoupling of the mechanical superposition mode from the optical mode by the destructive optomechanical interference. Overall, these studies establish that interference is an effective and essential tool for controlling the interactions between light and mechanical oscillators.

For the three-mode system shown in Fig. 1, two mechanical modes with frequencies $\omega_{m1}$ and $\omega_{m2}$, couple to an optical mode with frequency $\omega_0$, with the optomechanical coupling driven by two strong external laser fields, $E_1$ and $E_2$, which are respectively $\omega_{m1}$ and $\omega_{m2}$ below the optical resonance. The interaction Hamiltonian including only resonant processes is given by

$$V_R = \hbar \hat{a}^+ (e^{i\phi_1} G_1 \hat{b}_1 + e^{i\phi_2} G_2 \hat{b}_2) e^{i(\omega_s - \omega_0)t} + h.c. \qquad (1)$$

where $\hat{b}_1$ and $\hat{b}_2$ are the mechanical annihilation operators in their respective rotating frames, $\hat{a}$ is the annihilation operator for the optical mode in the rotating frame of a signal field with frequency $\omega_s$, $\phi_1$ and $\phi_2$ are the initial phases of $E_1$ and $E_2$, and $G_1$ and $G_2$ are the optomechanical coupling rates for the individual mechanical modes. Under these conditions, the mechanical system features bright and dark mechanical modes, described respectively by their annihilation operators,

$$\hat{b}_B = (e^{i\phi_1} G_1 \hat{b}_1 + e^{i\phi_2} G_2 \hat{b}_2)/\sqrt{G_1^2 + G_2^2}, \qquad (2a)$$

$$\hat{b}_D = (e^{-i\phi_2} G_2 \hat{b}_1 - e^{-i\phi_1} G_1 \hat{b}_2)/\sqrt{G_1^2 + G_2^2}. \qquad (2b)$$

With $G_1=G_2$, the two superposition modes in Eq. 2 are completely controlled by the relative optical phase, $\Delta\phi=\phi_2-\phi_1$. In particular, by making a $\pi$ phase shift in $\Delta\phi$, we can turn a bright mechanical mode into a dark mechanical mode. Our experimental studies not only demonstrate the constructive and destructive interferences underlying the bright and dark modes, respectively, but also show that the destructive interference suppresses optically-induced mechanical damping. These interference experiments have been carried out at room temperature and above the thermal background. They can also be extended to the quantum regime.

A silica microsphere with a diameter near 200 μm is used as a model multimode system. For our experiments, two mechanical whispering gallery (WG) modes, with frequencies $\omega_{m1}/2\pi=69.48$ MHz and $\omega_{m2}/2\pi=69.66$ MHz and damping rates $\gamma_1/2\pi=3.5$ kHz and $\gamma_2/2\pi=3.6$ kHz, are coupled to a WG optical resonance with a wavelength near 1.55 μm and with damping rate $\kappa/2\pi=1.6$ MHz. The optomechanical interactions take place via anti-Stokes Brillouin scattering of the optical driving fields from the mechanical modes[28-30]. The input optical power used for the weak signal field near the optical resonance is less than 0.01 mW. For the optical driving fields, the input optical powers used range from 0.6 to 1.2 mW. Additional experimental details including the experimental setup are presented in the supplementary materials[31].

We have developed a phase-dependent excitation-coupling approach to probe optomechanical interactions and especially interference processes. We first illustrate this approach using a two-mode system. As shown in the inset of Fig. 2, a weak optical signal field, $E_s$, with frequency $\omega_s=\omega_0$, and an optical driving field, $E_1$, with frequency $\omega_1=\omega_s-\omega_{m1}$, couple to the mechanical mode, converting the signal field in the optical mode to a mechanical excitation[32-34]. After $E_s$ is switched off, $E_1$ couples to the induced mechanical excitation, converting the mechanical excitation back to optical fields. In addition, we introduce a phase slip in $E_1$ right after $E_s$ is switched off. The initial phase of $E_1$ in the excitation stage is $\theta_1$. The phase is then changed to $\phi_1$ in the coupling stage (see Fig. 2).

Heterodyne-detected emissions from the optical mode, with $E_1$ as the local oscillator, are plotted in Fig. 2 as a function of time. The exponential decay of the emission following the leading edge of the signal pulse corresponds to the increasing conversion of the signal field in the optical mode into the mechanical excitation. The decay time, which sets the timescale for the excitation to reach steady state, is given by $1/[(1+C_1)\gamma_1]$, where $C_1 = 4G_1^2/\gamma_1\kappa = 1.6$ is the cooperativity for the optomechanical coupling. The decrease in the emission from the optical mode in the steady state shown in Fig. 2 corresponds to the dip in the spectral domain optomechanically-induced transparency (OMIT) experiment[34]. The exponential decay after $E_s$ is switched off corresponds to the conversion of the induced mechanical excitation back into optical fields. With $\kappa \gg (\gamma_1, G_1)$, this conversion process leads to irreversible optically-induced damping of the mechanical excitation, with the total damping rate given by $(1+C_1)\gamma_1$, as confirmed in Fig. 2. Note that interference also plays an important role in two-mode systems through OMIT[35]. However, the underlying optomechanical coupling cannot be controlled via a phase shift in the optical or mechanical excitations. The experimental result for the two-mode system shown in Fig. 2 is independent of $\theta_1$ as well as the phase slip $\phi_1-\theta_1$.

We now extend this approach to the three-mode system, for which two optical driving fields, $E_1$ and $E_2$, with frequencies $\omega_1=\omega_s-\omega_{m1}$ and $\omega_2=\omega_s-\omega_{m2}$, couple to the two mechanical modes. The pulse sequence of the experiment is shown in Fig. 3a. For simplicity, no phase slip is introduced for $E_2$, (i.e., $\theta_2=\phi_2$). In the coupling stage, the induced mechanical excitation is in a bright mechanical mode when $\phi_1=\theta_1$. The same excitation, however, is expected to be in a dark mechanical mode when $G_1=G_2$ and $\theta_1$ is $\pi$ out of phase with $\phi_1$. In general, the mechanical excitation can be a combination of both bright and dark modes.

Heterodyne-detected emissions from the optical mode are shown in Figs. 3b and 3c as a function of time. A spectral filter is used such that only the heterodyne beat at frequency $\omega_{m1}$, with either $E_1$ or $E_2$ as the local oscillator, is detected. The emissions in Figure 3b are detected during the excitation stage of the experiment. Similar to Fig. 2, the decay of the emission in Fig. 3b corresponds to the increasing conversion of the signal field in the optical mode into the mechanical excitations and shows an effective cooperativity of $C=1.4$. The emissions in Fig. 3c are obtained when $E_s$ is switched off. In this case, the optical driving fields convert the mechanical excitations back to optical fields, leading to optically-induced mechanical damping. As revealed in Fig. 3c, the optomechanical coupling process depends strongly on the phase slip $\phi_1$-$\theta_1$.

The heterodyne-detected optical emission energy obtained in a time span of 0.4 ms after $E_s$ is switched off is plotted in Fig. 3d as a function of $\phi_1$. These data are derived from experiments similar to those in Fig. 3c. The interference fringes observed in Fig. 3d are sinusoidal with a period of $2\pi$. The minima and maxima in the oscillations correspond respectively to the dark and bright mechanical modes. The sinusoidal oscillations reflect the switching of the mechanical system between the dark and bright modes as $\phi_1$ is varied. Similar oscillations are also observed when the heterodyne beat at frequency $\omega_{m2}$ is detected.

The interference underlying the oscillations shown in Fig. 3d takes place between two optomechanical coupling processes: red sideband coupling of $E_1$ to mechanical mode 1 and that of $E_2$ to mechanical mode 2. Under the condition of two-photon resonance, $\omega_1 + \omega_{m1} = \omega_2 + \omega_{m2}$, both processes result in converted optical fields at the same frequency. For the dark mechanical mode, the two converted fields interfere destructively; whereas for the bright mechanical mode, the two converted fields interfere constructively.

Destructive optomechanical interference not only leads to corresponding destructive interference in the converted optical fields, but also suppresses optically-induced mechanical damping. In a simple picture, the destructive interference prevents the conversion of the mechanical excitation to optical fields, thus inhibiting the optically-induced mechanical damping and effectively decoupling the mechanical system from the optical mode. To determine the damping rate of the dark mode, we append a measurement stage to the pulse sequence in Fig. 3a. As shown in Fig. 4a, after keeping the mechanical system in the dark mode for a duration of $\tau$, we switch the initial phase of $E_1$ back to $\theta_1$. Correspondingly, the mechanical system is switched back to the bright mode. Heterodyne-detected optical emissions occurring in the measurement stage probes directly the amplitude of the dark mode at the end of the coupling stage. The emission energy obtained for a time span of 0.4 ms in the measurement stage is plotted in Fig. 4b as function of $\tau$. Similar to Fig. 3, only the heterodyne beat at frequency $\omega_{m1}$ is detected. Note that the damping rate of the bright mechanical mode can be derived from experiments similar to those in Fig. 3c, in which we measure directly the heterodyne-detected optical emission as a function of time after $E_s$ is switched off.

The damping rate for the dark mechanical mode, derived from Fig. 4b, is $\gamma_D/2\pi$=7.8 kHz. In comparison, the damping rate for the bright mechanical mode obtained under otherwise the same experimental condition is $\gamma_B/2\pi$=11 kHz (see the inset of Fig. 4b), corresponding to $C$=2.1. The relative reduction in the optically-induced mechanical damping rate due to the destructive interference is $(\gamma_B - \gamma_D)/(\gamma_B - \bar{\gamma}) = 43\%$, where $\bar{\gamma} = (\gamma_1 + \gamma_2)/2$. The suppression is not complete, in part due to the slightly unequal damping rates of the two mechanical modes, and to a larger part due to optomechanical coupling processes that are not two-photon resonant. These processes include the coupling of $E_1$ to mechanical mode 2 and the coupling of $E_2$ to mechanical mode 1[31]. These two processes do not experience destructive interference, leading to effective damping of the mechanical modes.

For a theoretical analysis of the experimental results, we have used the semi-classical coupled-oscillator equations, with the equations of motion given by

$$\dot{\beta}_1 = -(\gamma_1/2)\beta_1 - ie^{-i\delta t - i\phi_1} G_1 \alpha \tag{3a}$$

$$\dot{\beta}_2 = -(\gamma_2/2)\beta_2 - ie^{-i\delta t - i\phi_2} G_2 \alpha \tag{3b}$$

$$\dot{\alpha} = -(i\Delta + \kappa/2)\alpha - i(e^{i\phi_1} G_1 \beta_1 + e^{i\phi_2} G_2 \beta_2)e^{i\delta t} + \sqrt{\kappa^{ext}} A_s \tag{3c}$$

where $\beta_1 = <\hat{b}_1>$, $\beta_2 = <\hat{b}_2>$, $\alpha = <\hat{a}>$, $\Delta = \omega_0 - \omega_s$, and $\kappa^{ext}$ is the cavity decay rate due to input-output coupling. The amplitude of the input signal field, $A_s$, is normalized such that $I_s = |A_s|^2$ is the photon flux. For simplicity, the above equations have assumed that the two-photon resonant condition is satisfied, with $\delta = \omega_s - \omega_1 - \omega_{m1} = \omega_s - \omega_2 - \omega_{m1}$, and have omitted coupling terms that are not two-photon resonant (the general equations are given in [31]). It is straight forward to show from Eq. 3 that with $\gamma_1 = \gamma_2$, the amplitude of the dark mode, $\beta_D = \langle \hat{b}_D \rangle$, is completely decoupled from the field in the optical mode. Theoretical calculations, which include both two-photon resonant and non-resonant optomechanical couplings and use experimentally determined parameters, are in good agreement with the experimental results on the fringe visibility shown in Fig. 3d and on the damping rate of the dark mode shown in Fig. 4b. As shown in Fig. 4b, the theoretical calculation that includes only two-photon resonant optomechanical coupling yields a damping rate for the dark mode, $\gamma/2\pi = 3.6$ kHz, nearly the same as $\gamma_1/2\pi$ and $\gamma_2/2\pi$. In this regard, the residual optically-induced mechanical damping for the dark mode is almost entirely due to the two-photon non-resonant couplings, which can be suppressed if the frequency separation between the two mechanical modes far exceed the optical linewidth.

In conclusion, our experimental studies have demonstrated that by varying the relative phase of the optical driving fields, we can directly access and manipulate optomechanical interference processes and thereby control optomechanical interactions in multimode optomechanical systems. Like its counterpart in multi-level or multi-qubit systems, optomechanical interferences play an essential role in the exploration and application of interactions between light and mechanical systems.

**Acknowledgements**

This work is supported by the National Science Foundation under grant No. 1606227.

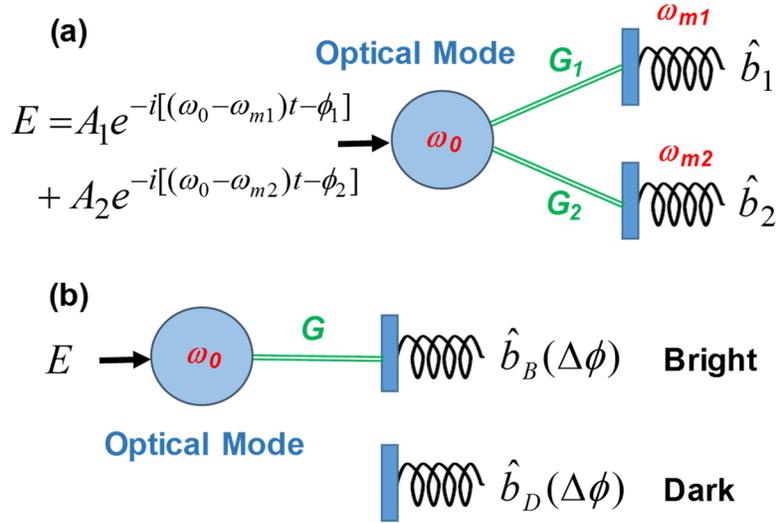

**Figure 1** (a) Schematic of a multimode optomechanical system driven by two optical fields via respective red sideband couplings. (b) Interference between the two optomechanical coupling processes leads to the formation of dark and bright mechanical modes that depend on the relative phase of the optical driving fields, $\Delta\phi=\phi_2-\phi_1$, with the dark mode decoupled from the optical mode.

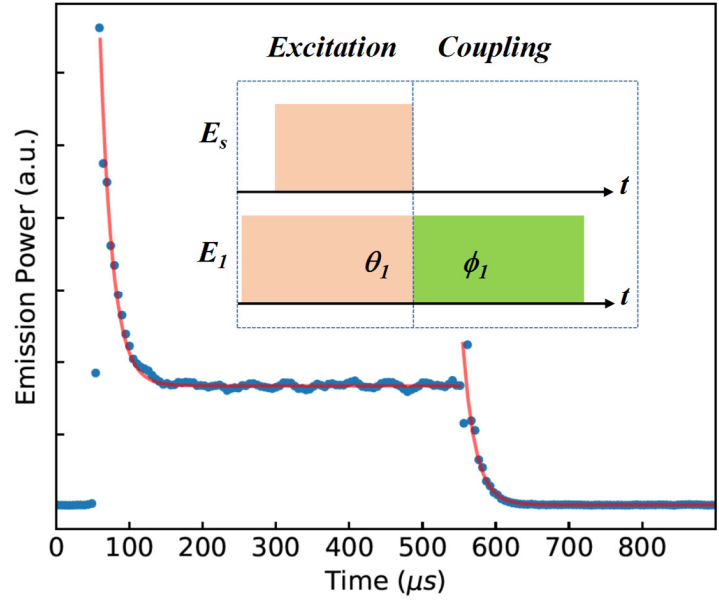

**Figure 2** Heterodyne-detected emissions (the dots) from the optical mode as a function of time in the two-mode system. Solid lines are numerical fits to single exponential decays with a decay rate, $\gamma/2\pi$=9.1 kHz. The first decay corresponds to the increasing conversion of $E_s$ to a mechanical excitation. The second decay corresponds to the conversion of the mechanical excitation to optical fields and the resulting mechanical damping, after $E_s$ is switched off. The inset shows the optical pulse sequence used, with $E_s$ (0.5 ms in duration) at $\omega_0$ and $E_1$ at the red sideband of $E_s$.

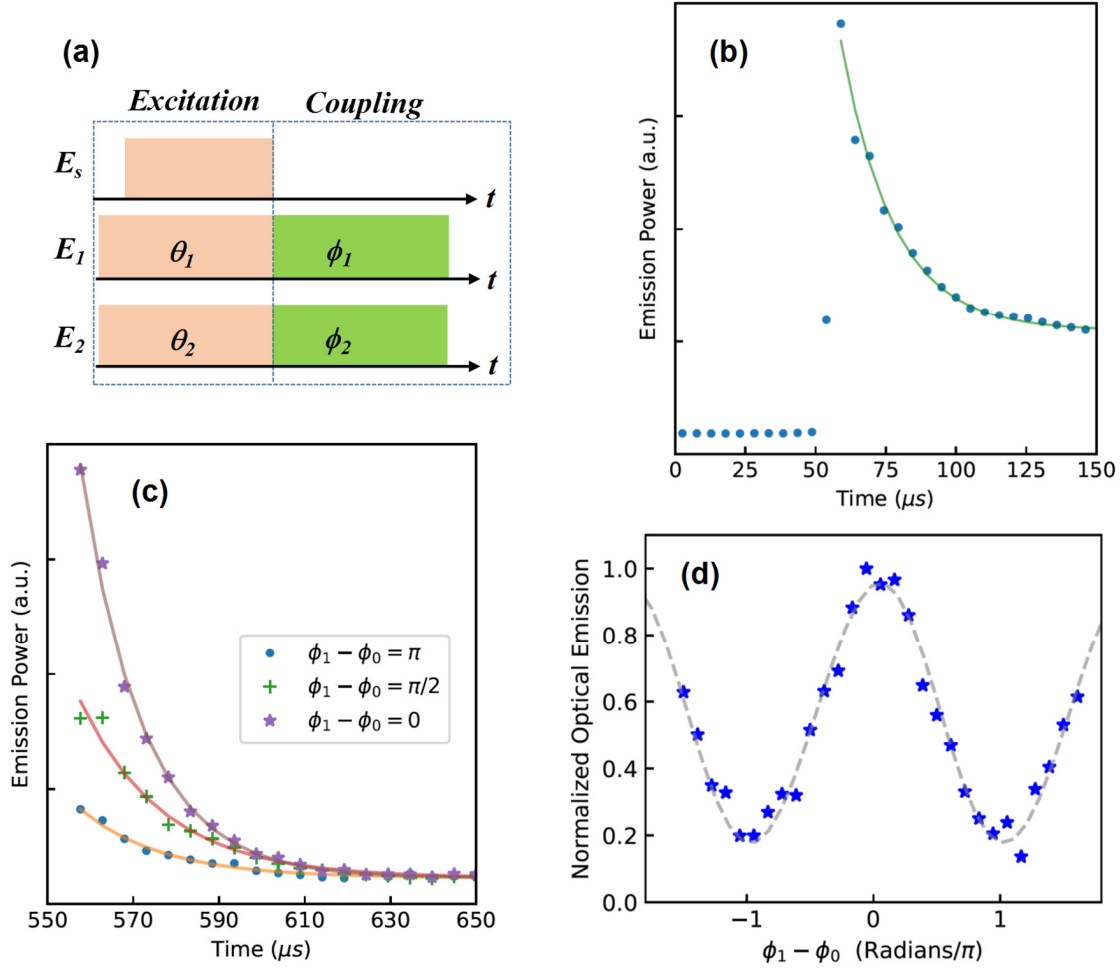

**Figure 3** (a) Optical pulse sequence used for optomechanical interference, with $E_s$ (0.5 ms in duration) at $\omega_0$ and $E_1$ and $E_2$ at the respective red sidebands of $E_s$. (b) Heterodyne-detected emissions from the optical mode as a function of time with $\theta_2=\phi_2$, when $E_s$ is on. (c) Heterodyne-detected emissions from the optical mode as a function of time at various $\phi_1$ with $\theta_2=\phi_2$, $C_1=1.3$, and $C_2=1$, when $E_s$ is off. Solid lines in (b) and (c) are numerical fits to single exponential decays. (d) The emission energy from the optical mode as a function of $\phi_1$, obtained in a time span of 0.4 ms after $E_s$ is switched off. The dashed line shows the theoretical calculation discussed in the text. $\phi_0$ is an offset such that the dark mode occurs when $\phi_1-\phi_0=\pi$.

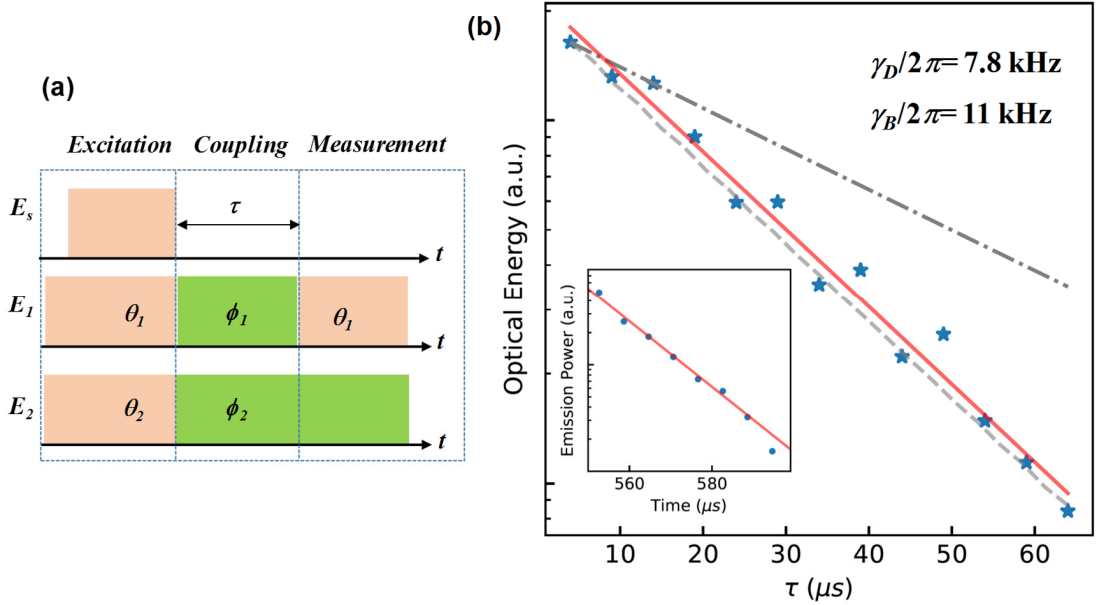

**Figure 4** (a) Optical pulse sequence used to probe the suppression of optically-induced mechanical damping due to destructive interference. (b) Heterodyne-detected optical emissions (the stars) from the optical mode obtained in the measurement stage as a function of $\tau$, with $\theta_2=\phi_2$ and $G_1=G_2$ and with $\phi_1$ adjusted such that the mechanical system is in the dark mode in the coupling stage. The dashed line shows the corresponding theoretical calculation discussed in the text. The dash-dotted line shows the theoretical calculation that includes only two-photon resonant optomechanical coupling, yielding a damping rate, $\gamma/2\pi=3.6$ kHz. The inset shows the measurement of the bright mode decay in the coupling stage. The solid lines are numerical fits of the experimental data to single exponential decays.